\documentclass[a4paper,twoside,10pt,fullpage]{article}
\newcommand{\affil}[1]{$^{\rm #1}$}
\usepackage[authoryear]{natbib}
\bibpunct{(}{)}{;}{n}{}{,}
\usepackage{graphicx}

\date{} 

\title{\large\bf\flushleft An all-sky Atlas of Radio/X-ray Associations}
\author{\parbox{\textwidth}{\flushleft
\vspace{-0.5cm}
{\it E. Flesch\affil{A,B}}\\
\vspace{0.4cm}
{\small \affil{A}\,P.O. Box 12520, Wellington, New Zealand}\\
{\small \affil{B}\,Email: eric@flesch.org}}}

\begin{document}
\twocolumn[
\begin{changemargin}{.8cm}{.5cm}
\begin{minipage}{.9\textwidth}
\vspace{-1cm}
\maketitle
\small{\bf Abstract:  An all-sky comprehensive catalogue of calculated radio and X-ray associations to optical objects is presented.  Included are X-ray sources from {\it XMM-Newton}, {\it Chandra} and {\it ROSAT} catalogues, radio sources from NVSS, FIRST and SUMSS catalogues, and optical data, identifications and redshifts from the APM, USNO-A, SDSS-DR7 and the extant literature. This `Atlas of Radio/X-ray Associations' inherits many techniques from the predecessor Quasars.org (2004) catalogue, but object selection is changed and processing tweaked.  Optical objects presented are those which are calculated with $\ge 40$\% confidence to be associated with radio/X-ray detections, totalling 602 570 objects in all, including 23 681 double radio lobe detections.  For each of these optical objects I display the calculated percentage probabilities of its being a QSO, galaxy, star, or erroneous radio/X-ray association, plus any identification from the literature.  The catalogue includes 105568 uninvestigated objects listed as 40\% to $>99$\% likely to be a QSO.  The catalogue is available at http://quasars.org/arxa.htm .}


\medskip
\medskip
\end{minipage}
\end{changemargin}
]
\small

\section{Introduction}

In 2004 we published the all-sky Quasars.org catalogue (QORG: Flesch \& Hardcastle 2004) as a compendium of all radio/X-ray associations and quasars, mapped onto an optical background using a uniform matching algorithm.  This was meant as a comprehensive resource for investigators, but included X-ray data only from {\it ROSAT} catalogues.  Since then, large X-ray source catalogues have been released from {\it XMM-Newton} and {\it Chandra} satellite operations, and large radio catalogues have been completed or amended.  Also, large new optical identification surveys have been completed, most notably the Sloan Digital Sky Legacy Survey (SDSS: Abazajian et al. 2009), and many smaller surveys.  These developments have left the QORG catalogue slightly dated; also I have refined the percentage calculations and improved the duplications handling.

I have accordingly included all new data to mid-2009 and updated the matching algorithm.  Also two significant changes in presentation are in order: (1) QORG endeavoured to present all quasars as at 2003; however, I do not wish to echo the large recent releases of new quasars from surveys like SDSS.  Therefore I constrain the object selection to only those calculated as having radio/X-ray associations.  (2) QORG presented only objects successfully mapped onto an optical background derived from Cambridge Automatic Plate Measuring machine (APM: McMahon \& Irwin 1992) and United States Naval Observatory (USNO-A: Monet 1998) data.  This seems too restricting now, given that SDSS and other surveys publish entirely suitable optical data.  Therefore I augment this background with all published objects and their accompanying optical data, taking care to avoid duplications.  This new presentation constitutes a change in scope from QORG.

I thus present a new comprehensive all-sky catalogue, the `Atlas of Radio/X-ray Associations' (hereafter: ARXA), which uses all good-resolution radio and X-ray source catalogs to mid-2009, and causally maps those sources onto all available optical data using techniques from QORG, with some enhancements.  Optical objects presented are those calculated with $\ge 40$\% confidence to be associated with radio/X-ray detections; the 40\% threshold is chosen as a lean towards completeness.  These optical objects total 602 570 in all, including 190 393 objects bearing identifications from the literature, 294 713 non-identified objects previously reported in the QORG catalogue (but here recalculated), and 117 320 objects reported here for the first time.  Using the individual confidences of association as expectation values, and taking objects identified from the literature as correctly associated, this catalogue should yield 500 286 correct associations of radio/X-ray detections to optical objects.  The individual confidence of associations are divided into percentage probabilities of the optical object being a QSO, galaxy or star.  Accordingly, this catalogue displays 105~568 uninvestigated objects listed as 40\% to $>99$\% likely to be a QSO; 72 522 of these were first presented in QORG, and 33 046 are newly presented here in ARXA.  This catalogue is intended as a bulk resource for investigators, and is obtainable from http://quasars.org/arxa.htm .  

In this paper I give an overview of ARXA input data, focusing on data newly added since QORG, and I describe refinements in the processing, and present the resultant ARXA catalogue.  For input data carried over from QORG, readers are asked to consult the QORG paper, which also gives full details of the matching techniques in its appendix A.  
   
\subsection{X-ray data}

The input X-ray data is comprised of {\it XMM-Newton} (XMM) and {\it Chandra} catalogues published to 2009, plus the four {\it ROSAT} (ROentgen SATellite) catalogues included and documented in QORG, which are the All-Sky Survey (RASS: Voges et al. 2000), High Resolution Imager (HRI: Voges et al. 1999), Position Sensitive Proportional Counter (PSPC: Voges et al. 1999), and the WGACAT (WGA: White Giommi Angelini 1994) catalogues.  As XMM and Chandra data were not featured in QORG, I document them here. 

Three XMM input files are utilized, these are the Incremental Second XMM-Newton Serendipitous Source Catalogue (2XMMi: Watson et al. 2009), the XMM-Newton Slew Survey catalogue release 1.3 (XMMSL: Saxton et al. 2008), and the XAssist XMM master sourcelist v2009 (XAssist: Ptak \& Griffiths 2001).  I process each of these separately, keeping only those with fluxes greater than their uncertainty, matching to optical within the XMM positional uncertainty, and then combining them into one file; where more than one input files match to the same optical object, the highest-confidence match is kept.  ARXA identifies these three source catalogues with their prefixes 2XMM, XMMSL or XMMX, respectively.  The merged XMM data contribute 57 778 X-ray-optical associations to ARXA. 

Similarly, three Chandra input files are utilized, these are the Chandra Source Catalog v1.01 (CSC: http://cxc.harvard.edu/csc), the Chandra Multiwavelength Project X-ray Point Source Catalog (ChaMP2: Kim et al. 2007), and the XAssist Chandra master sourcelist v2009 (XAssist: Ptak \& Griffiths 2001). Processing is as with the XMM data.  ARXA identifies these three source catalogues with their prefixes CXO, CXOMP or CXOX, respectively.  The merged Chandra data contribute 32 951 X-ray-optical associations to ARXA. 
 
The numbers sourced from these six input X-ray catalogues are shown in Table 1.  The 3rd column counts associations found per input catalogue, and the 4th column counts which are used, as ARXA displays just one XMM and/or Chandra identification per association.  The 5th column counts those associations available from only one of the three XMM input catalogues, and similarly for Chandra.

\begin{table}[h]
\begin{center}
\scriptsize
\caption{Associations contributed by the six XMM and Chandra source files. }\label{table1}
\begin{tabular}{lrrrr}
\hline
           & \# unique  & Number  & Number  & Number\\
Source     & and useful & assoc's & assoc's & unique\\
catalogue  &  sources   &  found  & in ARXA & assoc's\\
\hline
   2XMMi       & 249916 &  54980  &  51781  &  44845\\
   XMM Slew    &   7357 &   1881  &   1689  &   1662\\
   XMMX        &  63404 &  11155  &   4308  &   1124\\
   All XMM     &        &         &  57778\\
   CSC / CXO   &  93685 &  25917  &  21271  &  13688\\
   ChaMP2      &   6512 &   1936  &    949  &    272\\
   CXOX        &  94103 &  18769  &  10731  &   6305\\
   All Chandra &        &         &  32951\\
\hline
\end{tabular}
\medskip\\
\end{center}
\end{table}

These input catalogues provide astrometry both raw and adjusted via an optical solution.  For consistency, I elect to apply our own optical solutions onto the raw source positions, using the probability-based technique documented in QORG.  40\% of the XMM sources are thus shifted 1 to 2 arcsec, with a few of 3 or 4 arcsec.  The Chandra astrometry fits strongly, and I find shifts of 1 arcsec for only 4\% of its sources, with a few outliers of 2 or 3 arcsec.  This represents an order of improvement over the {\it ROSAT} astrometry treated in the QORG paper.       

X-ray sources are usually core emissions, but in practice offsets result from extended emission and the astrometric imprecision inherent in reducing small counts.  The input sources are listed with positional uncertainties, and we accept optical selections only within those uncertainties.  However, a few farther-offset associations are accepted in the course of performing de-duplications.  Table 2 shows the counts, by astrometric offset of X-ray source from optical.    

\begin{table}
\begin{center}
\scriptsize
\caption{XMM and Chandra sourced associations in ARXA, by astrometric offset to optical. }\label{table2}
\begin{tabular}{rrrrr}
\hline
Astro-  &         & Mean         &         & Mean\\
metric  &  No. of & confidence   & No. of  & confidence\\
offset  &  XMM    & of XMM       & Chandra & of Chandra\\
(arcsec) & assoc's & assoc's (\%) & assoc's & assoc's (\%)\\
\hline
     0  &    8088 &  97.5 &   7627 &  98.3\\
     1  &   24752 &  91.9 &  16478 &  92.1\\
     2  &   16103 &  85.0 &   6451 &  84.8\\
     3  &    8479 &  76.3 &   1777 &  75.4\\
     4  &     302 &  75.2 &    608 &  69.0\\
     5  &      11 &  74.1 &      4 &  56.5\\
     6  &      11 &  73.5 &      5 &  79.4\\
     7  &      15 &  75.0 &      1 &  60.0\\
     8  &      10 &  59.9 &        &      \\
     9  &       4 &  74.5 &        &      \\
    10  &       4 &  78.5 &        &      \\
    11  &       1 &  60.0 &        &      \\
 total  &   57778 &  88.4 &  32951 &  90.8\\
\hline
\end{tabular}
\medskip\\
\end{center}
\end{table}

Inspection of the large-offset outliers of Table 2 show these are mostly large galaxies or bright stars, which often feature multiple signatures.  One example is CXO J093552.7+612112 which is shown as associated to UGC 5101 at an offset of 6 arcsec.  Also associated to UGC 5101 are XMMX J093551.7+612112, 2RXP J093551.8+612110 and FIRST J093551.6+612111, all at offsets of 0--2 arcsec.  Investigation shows there is another Chandra source CXO J093551.6+612111 which aligns excellently with the others, but CXO J093552.7+612112 was also retained by the processing because it is precisely co-positioned onto a component of UGC 5101 which was deconvolved in our background USNO-A data; subsequent de-duplication retained just the one Chandra association with the highest confidence of association.  Thus these outliers stem from such artefacts typical of large data, but their inclusion is valid.

\subsection{Radio data}

The radio data is comprised of the large NRAO VLA Sky Survey catalog (NVSS: Condon et al. 1998), the re-release (2008) of the Faint Images of the Radio Sky at Twenty-cm survey catalog (FIRST: White et al. 1997), and the completed Sydney University Molonglo Sky Survey (SUMSS: Murphy et al. 2007).  SUMSS is taken to include the Molonglo Galactic Plane Survey (MGPS-2: same attribution as SUMSS), and of the 61 659 SUMSS sources displayed in ARXA, 3954 carry the MGPS identifier.  Core associations are presented from all three source catalogues, which in practice come to at most two core associations per object because FIRST and SUMSS do not overlap.  Double lobe associations are also presented, but just one set per optical object, for brevity.

Processing is as detailed in the QORG paper.  The 2008 release of the FIRST catalogue has additions and tweaks compared with its 2003 version, and now displays a sidelobe probability figure.  Frequency analysis and FIRST cutout inspection led me to adopt a cutoff of 66.7\% sidelobe probability -- sources more likely to be sidelobes were dropped, and retained sources with high sidelobe probability are less likely to match to optical objects anyway.  

Total counts of these radio and X-ray input catalogues and the ARXA associations derived from them are in Table 3.

\begin{table}
\begin{center}
\scriptsize
\caption{Radio/X-ray Associations presented in the ARXA.}\label{table3}
\begin{tabular}{lrrr}
\hline
          & \# distinct & \# core & \# double\\
Source    & astrometric & assoc's & lobes\\
catalogue & sources     & in ARXA & in ARXA\\
\hline
   FIRST     &    792777  &   173383  &  12844\\
   NVSS      &   1810664  &   266148  &   8308\\
   SUMSS     &    259896  &    59138  &   2529\\
   All radio &            &   417075  &  23681\\
   XMM       &    313015  &    57778\\
   Chandra   &    183494  &    32951\\
   HRI       &     56398  &    15523\\
   RASS      &    124730  &    47486\\
   PSPC      &    102005  &    35607\\
   WGA       &     88578  &    24226\\
   All X-ray &            &   174337\\
\hline
\end{tabular}
\medskip\\
\end{center}
\end{table}

\subsection{Identifications data}

ARXA is intended as a comprehensive resource for investigators, so it behoves us to identify its optical objects from the literature where possible.  All input identification catalogues cited in the QORG paper are included, some having since had large updates (most notably the SDSS data releases 2 through 7), and also newly published spectroscopic and photometric catalogues, and numerous small surveys.  ARXA displays citations for the name and redshift of each optical object, and a full listing of those citations and respective catalogues is in the ReadMe  (http://quasars.org/arxa/ARXA-ReadMe.txt).

The primary catalogue used for identification of QSOs, AGN and BL Lacs is the Catalogue of Quasars and Active Nuclei, 12th edition (Veron: V\'eron-Cetty \& V\'eron 2006).  The Veron catalogue uses an absolute-magnitude threshold to differentiate a QSO classification from an AGN classification, and ARXA adheres to that.  QSO identifications are also taken from the NASA/IPAC Extragalactic Database (NED), although deduplication must be scrupulously done because of astrometric mismatches between the Veron and NED data.  These two catalogues provide historical names.  Also included are the many large and small QSO releases to 2009, notably SDSS-DR7, the 2dF-SDSS LRG and QSO catalogue (2SLAQ: Croom et al. 2009), and the 2dF QSO Redshift Survey (2QZ: Croom et al. 2004).  The total counts found to be associated to radio/X-ray sources are 21 735 catalogued QSOs, 6941 AGN and 803 BL Lacs.  15 241 of these objects stem from the SDSS survey.  There are also 24 717 SDSS-based photometric quasars (NBCKDE: Richards et al. 2009).

Our primary source for galaxy identification is the Principal Galaxy Catalogue (PGC: Paturel et al. 2003), which has 983 213 objects and historical names.  The SDSS-DR7 makes spectroscopic identification of 871 054 galaxies, the 2dF Galaxy Redshift Survey (2dFGS: Colless et al. 2001) has 236 078 galaxies, and the 6dF Galaxy Survey final release (6dFGS: Jones et al. 2009) has 124 575 galaxies.  Other galaxy catalogues are documented in the QORG paper and ARXA ReadMe.  The total count presented in ARXA as associated to radio/X-ray detections, is 90 006 catalogued galaxies and 31 658 SDSS-based photometric red galaxies (MegaZ-LRG: Abdalla et al. 2008).  Note that some nearby galaxies known to be radio/X-ray emitters are missing from ARXA because our astrometric matching is scaled to objects not so large on the sky.

The remaining identification is that of star.  Many large surveys have released star identifications, and I use the Tycho catalogue (Hog et al. 2000) as a bright star identifier although its objects are formally just point sources; the Lick North Proper Motion catalog (NPM: Klemola et al. 1987) is also well represented.  See the QORG paper for discussion of stellar identification and listing of many of the input catalogues used.  The total count presented in ARXA as associated to radio/X-ray detections, is 13 844 stars, for which we have about 8x more X-ray associations than radio.  Only for fainter ($>17$ mag) stars are radio associations as common as X-ray.  Table 4 gives counts provided by the largest input identification catalogues.  

\begin{table}
\begin{center}
\scriptsize
\caption{Identified objects presented in the ARXA, by source catalogue.}\label{table4}
\begin{tabular}{lcrrr}
\hline
  & object & all & radio & X-ray\\
catalogue & type & IDs & associated & associated\\
\hline
   PGC        & galaxy & 54670  &  49627  &   7356\\
   SDSS       & all & 50517  &  41290  &  10537\\
   MegaZ-LRG  & galaxy & 31658  &  30459  &   1345\\
   NBCKDE     & QSO & 24717  &  12475  &  12709\\
   TYCHO      & star &  8646  &    489  &   8192\\
   Veron      & QSO &  8482  &   4873  &   5432\\
   6dFGS      & galaxy &  3003  &   2330  &    842\\
   NPM        & star &  1767  &    129  &   1648\\
   2dFGS      & galaxy &  1431  &   1178  &    277\\
   2QZ        & QSO &  1175  &    502  &    695\\
   all others & all &  4327  &   1071  &   3469\\
\hline
\end{tabular}
\medskip\\
\end{center}
\end{table}

This leaves 412 177 radio/X-ray associated optical objects without identification from the literature and calculated as being $\ge 40$\% likely to be associated to radio/X-ray detections.  Of these, 294 713 associations were first presented in QORG, although recalculated here, and 117 320 are newly presented here in ARXA.  Of all these uninvestigated objects, 105 568 are listed as $\ge 40$\% likely to be a QSO, being 72 522 first presented in QORG and 33 046 newly presented here.  Table 5 gives counts of these uninvestigated objects, in total and also just for newly ARXA presented, binned by confidence of association and QSO probability. 
 
\begin{table}
\begin{center}
\scriptsize
\caption{Non-identified objects presented in the ARXA, all and new.}\label{table5}
\begin{tabular}{crrrr}
\hline
             &  all    &  all &   new   &  new\\
 probability & conf of &  QSO & conf of &  QSO\\
 percentage  & assoc'n & prob & assoc'n & prob\\
\hline
   40-49  &   57480 &  18662 &  32110 &  6735\\ 
   50-59  &   53716 &  15771 &  17461 &  4199\\
   60-69  &   46682 &  13682 &  10155 &  3286\\
   70-79  &   51748 &  16839 &   9968 &  3090\\
   80-89  &   65884 &  19725 &  13239 &  6218\\
   90-99+ &  136523 &  20889 &  34387 &  9518\\
   total  &  412033 & 105568 & 117320 & 33046\\
\hline
\end{tabular}
\medskip\\
\end{center}
\end{table}
 
\section{Processing issues and miscellaneous notes}

Our algorithm which calculates probability of association of radio/X-ray sources to optical objects is explained in full in the QORG paper and its appendix A.  In short, it consists of binning the optical objects by colour, psf, and astrometric offset to the radio/X-ray source, and then comparing the areal density of each bin compared with its average over all sky of the same overall object density.  Twice the average areal density means that we expect that half of such objects are causally associated, and so on.  Double lobe associations are calculated heuristically.  The QORG paper details processing issues; additional issues handled in the ARXA processing are described here.

Large input catalogues derived from surveys typically have some rows of lesser quality, whether flagged as such or missing some photometric information, or having fluxes less than their errors, etc.  I exclude these to keep false positives out of ARXA.  The two input photometric catalogues, NBCKDE and MegaZ-LRG, assign probabilities to their objects that they are indeed QSOs or galaxies.  Such probability should increase if an association to a radio/X-ray source is found.  Still, I wish to avoid displaying an erroneous identification as a QSO or galaxy, so I apply a cutoff threshold.  Numerical analysis shows that a probability cutoff of 60\% is suitable to accept NBCKDE QSO identifications, and 80\% for MegaZ-LRG galaxies.      

The Veron and NED QSO catalogues include data from older surveys with imprecise astrometry which have no obvious match to the optical data.  I have used a heuristic algorithm to determine the best optical objects for these.  Comparison to finding charts (where available) shows accuracy of about 90\%.  

The QORG paper, section A.6.1, describes a correction applied to those associations with large astrometric offsets ($>6$ arcsec) of the radio/X-ray source to optical.  We applied a linear-with-offset reduction in the probability of association, to deter false positives at large offsets.  It was an arbitrary solution to known issues at large offsets, such as increased likelihood of multiple optical candidates and true sources being too faint for our optical data.  We deliberately made the correction large, because, as we stated in the paper, `we feel it is more excusable to under-represent true far-offset associations than it is to over-represent false ones'.  Now with passage of 5 years, the subsequent publication of SDSS Data Releases 2 through 7 enables comparison of QORG optical selections for radio/X-ray sources with the new SDSS identifications.  This test is presented online at http://quasars.org/docs/Testing-QORG-via-SDSS-DR7.txt, and its first table is reproduced here as Table 6.  

This table shows that (1) optical identification accuracy exceeded nominal, and (2) accuracy was at least 75\%, even for low nominal confidence.  These low confidence bins are generally those of large astrometric offset, and shows that our arbitrary correction (for large-offset associations) was too great.  Analysis now has led me to apply a radio/X-ray survey specific solution, with on average about half the QORG correction.  This solution is designed to yield performance close to, but still exceeding, nominal confidence percentages at large astrometric offsets.  The effect is to include about 20 000 low-to-medium confidence associations which had been excluded from QORG.  Numbers of associations from {\it ROSAT} RASS are most increased by this, up about 50\%, because far-offset matching is needed to accomodate the poor astrometric precision of RASS.

\begin{table}
\begin{center}
\scriptsize
\caption{SDSS DR2--DR7 identifications compared against 36118 QORG associations}\label{table6}
\begin{tabular}{rrrr}
\hline
 QORG-listed  & \# hits; ie, & \# misses, ie, &  hit\\
  confidence  &  confirmed   &   SDSS says    &  pct\\
 pct (binned) &   by SDSS    & diff optical\\
\hline
   40-42  &     144  &    36  &  80.0\\
      45  &     265  &    84  &  75.9\\
      50  &     335  &   106  &  76.0\\
      55  &     382  &   100  &  79.3\\
      60  &     463  &    98  &  82.5\\
      65  &     502  &    94  &  84.2\\
      70  &     712  &   112  &  86.4\\
      75  &     749  &   118  &  86.4\\
      80  &    1041  &   144  &  87.8\\
      85  &    1561  &   130  &  92.3\\
      90  &    2476  &   176  &  93.4\\
      95  &    4713  &   215  &  95.6\\
  98-100  &   20929  &   433  &  98.0\\
   total  &   34272  &  1846  &  94.9\\
\hline
\end{tabular}
\medskip\\
\end{center}
\end{table}

L\'opez-Corredoira et al. (LGMGA: L\'opez-Corredoira et al. 2008) found that QORG QSO-probabilities underperformed for bright B$<17$ objects.  This is correct, as optical brightness was not one of our classifiers, and the optically brightest outliers of our identifications are naturally less likely to be QSOs than the mainstream.  However, any newly discovered bright QSO would be of particular interest, so I have not classified by optical brightness in ARXA either; the numbers involved are small anyway.  LGMGA also identified objects near ($<60$ arcsec) to bright galaxies as underperformed in QORG, but we find that the problem is caused by HII/starburst zones and deblending artefacts in the disks of bright galaxies which appear as bluish sources in our background optical APM/USNO-A data.  Thus the problem exists only as an artefact within the visible disks of bright galaxies.  These are difficult to identify from the data, and can be interesting anyway (e.g., the possibly background QSO on the disk of NGC 7314, J223546.2-260429), so I retain them in ARXA; however, I have included any identifications from LGMGA. 

Another artefact is a small number of nominal double lobe declarations in the Galactic dust lane.  These are unlikely to be true, but there is a residual chance that some interesting new or hidden background object could be pinpointed by this; thus I leave these in ARXA.

Processing to identify radio association has been done in isolation from the X-ray processing.  Thus optically near neighbours can be identified as a radio source on the one hand and an X-ray source on the other.  De-duplication will combine these only if the confidence of association is low for one of them.

De-duplication of optical objects was done as the last step, and the method is different to that used for QORG.  One type of duplication is when two optical data points represent the same true optical object.  Another type is when two optical objects have radio/X-ray associations belonging to just the one true optical source.  At large astrometric separations, two associations can be combined only if one is at very low confidence.  The heuristic rules were worked out with much testing against Digitized Sky Survey images and FIRST cutouts.  In total, 17,168 optical duplicates were removed.

\section{The Atlas of Radio/X-ray Associations}

The catalogue is available at http://quasars.org/arxa.htm, and is written as one line per optical object. The catalogue presents unique `best' associations, so optical objects and radio/X-ray sources appear once only.  The presentation is simplified from that of QORG: just the radio/X-ray association name is displayed, and the user can refer to the original radio/X-ray catalogue for information on that source.  The catalogue is 20Mb zipped, and Table 7 shows the first few lines. 

\begin{table*} 
\begin{center}
\scriptsize
\caption{Sample lines from the Atlas of Radio/X-ray Associations}\label{table7} 
\tiny 
\begin{tabular}{@{\hspace{0pt}}l@{\hspace{1pt}}l@{\hspace{1pt}}l@{\hspace{1pt}}c@{\hspace{1pt}}c@{\hspace{1pt}}l@{\hspace{1pt}}c@{\hspace{1pt}}c@{\hspace{1pt}}c@{\hspace{1pt}}c@{\hspace{1pt}}c@{\hspace{1pt}}r@{\hspace{1pt}}r@{\hspace{1pt}}c@{\hspace{1pt}}r@{\hspace{1pt}}l@{\hspace{1pt}}l@{\hspace{0pt}}} 
\\ 
\hline 
optical J2000 / & identification & obj & \multicolumn{2}{c}{optical mag} & opt & \multicolumn{2}{c}{PSF} & & \multicolumn{2}{c}{source} & \multicolumn{4}{c}{percentages} & 1st & 2nd\\
ARXA ID & from literature & type & red & blue & com & R & B & z & ID & z & qso & gal & str & err & association & association (etc)\\ 
\hline  
000000.6+321230 & PGC 1985872              & GRX & 12.2 & 16.0 & p  & 2 & 2 &       & PG &    &  0 &  90 & 2 &  8 & NVSS J000000.1+321233 & 1RXS J000001.3+321247\\
000000.7-083628 &                          & R   & 20.3 & 20.7 &    & 1 & - &       &    &    & 33 &   5 & 2 & 60 & FIRSTJ000000.1-083621\\  
000000.9-200447 &                          & RX  & 18.7 & 20.3 & pv & 1 & 1 &       & QO &    &  4 &  63 & 0 & 33 & NVSS J000000.2-200448 & 1RXS J000000.9-200444\\
000001.0+012416 &                          & R   & 20.3 &  0   &    & 1 & x &       & QO &    &  1 &  98 & 1 &  0 & FIRSTJ000001.0+012415\\
000001.3-020200 & FIRST J00000-0202        & QR  & 19.7 & 21.0 & p  & - & - & 1.356 & VE & VE & 85 &   7 & 6 &  2 & FIRSTJ000001.2-020200\\
000001.3-063114 &                          & R   & 18.6 & 21.1 &    & 1 & 1 &       & QO &    &  0 &  95 & 0 &  5 & NVSS J000001.4-063113\\
000001.4+111938 &                          & X   & 19.7 &  0   & p  & - & x &       &    &    & 19 &  15 & 9 & 57 &                       & 1RXS J000001.9+111948\\
000001.5-092940 & SDSS J000001.5-092940    & GR  & 17.3 & 19.8 & p  & 1 & 1 & 0.191 & SD & SD &  1 &  97 & 1 &  1 & NVSS J000001.4-092940 & FIRSTJ000001.5-092940\\
000001.6-251707 &                          & X   &  0   & 22.0 &    & x & - &       &    &    & 93 &   4 & 0 &  3 &                       & 2XMM J000001.6-251706\\ 
000001.6-420429 &                          & R   & 19.4 & 19.9 &    & - & - &       &    &    & 89 &   1 & 1 &  9 & SUMSSJ000001.7-420432\\ 
000001.8-094652 & NBCK J000001.88-094652.0 & qX  & 18.7 & 19.1 & p  & - & - & 1.000 & NB & NB & 93 &   1 & 1 &  5 &                       & CXO J000002.0-094649\\
000002.0-152435 &                          & R   & 16.7 & 19.1 & p  & 1 & 1 &       & QO &    &  1 &  92 & 0 &  7 & NVSS J000001.9-152435\\
000002.1+155254 & CGCG 456-13              & GR  & 11.1 & 11.4 & p  & 1 & 1 & 0.020 & PG & SD &  0 &  68 & 1 & 31 & NVSS J000001.6+155254\\
000002.1-093136 & SDSS J000002.1-093136    & GR  & 19.2 & 22.9 &    & 1 & 2 & 0.466 & SD & SD &  0 & 100 & 0 &  0 & FIRSTJ000002.1-093136\\  
000002.3-473423 &                          & R   & 20.4 & 22.5 &    & - & - &       &    &    & 12 &  31 & 1 & 56 & SUMSSJ000002.1-473429\\
000002.4+051717 & PGC 1281052              & GR  & 15.3 & 18.9 & p  & 2 & 2 &       & PG &    &  0 &  99 & 0 &  1 & NVSS J000002.3+051717\\
000002.4-042804 & IRASF 23574-0444         & GR  & 14.2 & 16.1 & p  & 1 & 1 & 0.102 & PG & 6d &  0 &  89 & 0 & 11 & NVSS J000002.3-042803\\
000002.5+395733 &                          & R   & 16.6 & 19.9 & p  & 2 & 2 &       & QO &    &  0 &  79 & 0 & 21 & NVSS J000003.3+395733\\ 
000002.6-321531 &                          & X   & 20.7 & 21.1 & p  & - & - &       & QO &    & 98 &   1 & 1 &  0 & 2XMM J000002.6-321530 & 2RXP J000002.8-321527\\
000002.7-251137 & XMM J00000-2511          & QX  &  0   & 21.8 &    & x & - & 1.314 & VE & VE & 93 &   4 & 0 &  3 &                       & 2XMM J000002.7-251136\\
000002.8-233911 & IRAS 23574-2356          & GR  & 16.9 & 18.2 & p  & 1 & 1 &       & PG &    &  8 &  79 & 1 & 12 & NVSS J000002.7-233910\\ 
\hline
\end{tabular} 
\medskip\\
\end{center}
\end{table*} 

Table 7 shows the data structure of the ARXA catalogue, although the right end is truncated to save space.  The ReadMe file is provided on site which gives full file layouts, field definitions and supporting information, plus citations for all input catalogues; I give only an overview here. Column 1 displays the optical coordinates
(epoch J2000) which doubles as the IAU-recommended name of the object, e.g., ARXA J040904.9-364744.  Column 2 gives the name of the object where it is identified from the literature.  This is left blank if previously identified only in QORG (column 10 = `QO'), in which case the QORG name is e.g. QORG J000001.0+012416, using the astrometry from column 1.  Column 3 summarizes any identification of, and associations with, the optical object: R=radio source, X=X-ray source, 2=double lobe declaration, Q=known quasar, A=AGN, G=galaxy, S=star, B=BL Lac object, q=photometric quasar, g=photometric galaxy.  Columns 4 and 5 give the red and blue magnitudes respectively, and column 6 flags if those magnitudes are POSS-I (='p') or other photometry, plus other comments.  Columns 7 and 8 give the point spread function
(psf) classification of the two colours: '-'=stellar, '1'=fuzzy, '2'=extended, 'n'=no psf and 'x'=object not seen in this colour. Column 9 gives the redshift, if known.  Column 10 gives the citation for the name in column 2 and/or object type in column 3.  The legend for these 2-byte citations is given in the ReadMe.  Column 11 gives the citation for the redshift.  Columns 12-15 give the calculated probability (independent of any identification from the literature) that the object is turn a QSO, galaxy, star, or erroneous radio/X-ray association.  Columns 16 and 17 here show radio/X-ray association names, but the actual ARXA structure is that column 16 is exclusively for NVSS associations, column 17 for FIRST/SUMSS associations (which don't overlap), and so on for XMM, RASS, PSPC, WGA, HRI and Chandra associations, and then the two signatures of a double radio lobe declaration. 

\section{Summary}

This paper presents the Atlas of Radio/X-ray Associations (ARXA), which is intended as grand compilation of the large good-resolution surveys of the radio and X-ray sky overlaid and aligned onto the optical sky.  It uses the completed {\it ROSAT}, NVSS, FIRST and SUMSS catalogues and {\it XMM-Newton} and {\it Chandra} data to 2009.  It provides calculated optical associations for these together with comprehensive identifications of known objects with the intention of presenting an informative map to help formulate and support pointed investigations.  Its counts are 602 570 associated optical objects in total, including 105 568 quasar candidates. 

\section*{Acknowledgements}

Sincerest thanks to Martin Hardcastle for advice and encouragement.

This research has made use of data obtained from the Chandra Source Catalog (ADS/Sa.CXO\#CSC 2009 v1.01), provided by the Chandra X-ray Center (CXC) as part of the Chandra Data Archive.  The National Radio Astronomy Observatory is a facility of the National Science Foundation operated under cooperative agreement by Associated Universities, Inc. The NASA/IPAC Extragalactic Database (NED: nedwww.ipac.caltech.edu) is operated by the Jet Propulsion Laboratory, California Institute of Technology, under contract with the National Aeronautics and Space Administration.  The Digitized Sky Surveys were produced at the Space Telescope Science Institute under U.S. Government grant NAG W-2166.

\end{document}